# MOF-Based Polymeric Nanocomposite Films as Potential Materials for Drug Delivery Devices in Ocular Therapeutics

J. Gandara-Loe,[1,ϕ] B.E. Souza,[2,ϕ] A. Missyul,[3] G. Giraldo,[4] J.-C. Tan,[2] J. Silvestre-Albero[1,*]

[1]Laboratorio de Materiales Avanzados, Departamento de Química Inorgánica-IUMA, Universidad de Alicante, E-03690 San Vicente del Raspeig, Spain

[2]Multifunctional Materials & Composites (MMC) Laboratory, Department of Engineering Science, University of Oxford, Parks Road, Oxford OX1 3PJ, UK

[3] CELLS-ALBA Synchrotron, E-08290, Cerdanyola del Vallés, Spain

[4] Clínica Clofan, Carrera 48 # 19 A 40, Medellín, Colombia.

ϕ These two authors contributed equally.

*Abstract*

Novel MOF-based polymer nanocomposite films were successfully prepared using Zr-based UiO-67 as a metal-organic framework (MOF) and polyurethane (PU) as a polymeric matrix. Synchrotron X-ray powder diffraction (SXRPD) analysis confirms the improved stability of the UiO-67 embedded nanocrystals and scanning electron microscopy images confirm their homogeneous distribution (average crystal size ~ 100-200 nm) within the 50-μm thick film. Accessibility to the inner porous structure of the embedded MOFs was completely suppressed for $N_2$ at cryogenic temperatures. However, ethylene adsorption measurements at 25ºC confirm that at least 45% of the MOF crystals are fully accessible for gas phase adsorption of non-polar molecules. Although this partial blockage limits the adsorption performance of the embedded MOFs for ocular drugs (e.g., brimonidine tartrate) compared to the pure MOF, an almost 60-fold improvement in the adsorption capacity was observed for PU matrix after incorporation of the UiO-67 nanocrystals. UiO-67@PU nanocomposite exhibits a prolonged release of brimonidine (up to 14 days were quantified). Finally, the combined use of SXRPD, thermogravimetric analysis (TGA) and FTIR analysis confirmed the presence of the drug in the nanocomposite film, the stability of the MOF framework and the drug upon loading, and the presence of brimonidine in an amorphous phase once adsorbed. These results open the gate towards the application of these polymeric nanocomposite films for drug delivery in optical therapeutics, either as a component of contact lens, in the composition of lacrimal stoppers (e.g., punctal plugs) or in sub-tenon inserts.

**Supporting Information.** Physicochemical characterization of the synthesized samples (synchrotron X-ray powder diffraction, TGA profiles and gas adsorption measurements), adsorption kinetics for brimonidine and mathematical models applied are included in the Supporting Information.

## 1. Introduction

Glaucoma is a pathological eye disorder associated with an increase in the intraocular pressure (IOP) and one of the leading causes of irreversible blindness worldwide.[1] Approximately, 70 millions middle-aged people and elderly are affected by its common form, open angle glaucoma whereof 10% ends in bilateral blindness.[2] Among the different drugs to treat glaucoma, brimonidine tartrate is one of the most widely applied. Brimonidine is an alpha-adrenergic agonist able to reduce the ocular pressure through the constriction in the blood vessels, ending in the decrease of the aqueous humour production.[3]

Conventional drug delivery systems such as eye droplets represent 90% of the marketed ophthalmic formulations.[4,5] However, severe constrains are associated with this topical approach such as tear turnover, fast nasolacrimal drainage and reflex blinking, thus ending in a non-optimal dosage.[6] Roughly, only 5% of the drug applied topically reaches the deeper ocular tissues, thus forcing pharmaceutical producers to increase the drug concentration, with the associated increase in the toxicity and, indirectly, the risk of side effects.[7] Another limitation of these topical administration routes is the low compliance of patients, mainly elderly, to strictly follow the administration protocol (administration of a number of droplets several times per day).

The development of more efficient ocular drug delivery systems with well-designed and prolonged release kinetics remains a challenge in materials science and ophthalmology. Nanocarriers such as polyacrylic acid nanoparticles,[8] chitosan nanoparticles,[9] nano-vesicles,[10] and layered double hydroxides (LDH)[11] have been reported as promising alternatives for topical brimonidine dosage. However, the main limitation of some of these materials for potential application falls in the physical (low gravimetric capacity for the drug) and textural properties.

Novel drug administration platforms to treat ocular disorders prepared from polymeric materials (solid or semi-solid inserts) have gained a large popularity in the last few years.[12–14] The potential advantage of these polymeric devices is the accurate dosing, increased ocular residence time, reduction of systemic side effects or better patient compliance, just to mention some.[15] Due to the potential of these devices in ocular drug delivery, several companies have patented and commercialized them. For instance, one of the first marketed ocular insert has been commercialized by Alza (Vacville, CA) as Ocusert® which are used to dose anti-glaucoma drug pilocarpine for a maximum of 5-7 days.[16,17] Although these are excellent numbers, the absence of a well-defined regular 3D network within these polymeric matrices limits their total drug uptake and hinders a controlled release.

Based on these premises, the design of novel functional ocular polymeric devices through the incorporation of perfectly designed high-capacity nanofillers would be a key stepping stone to increase the versatility and impact of these inserts in nanomedicine. A potential approach not widely explored in the literature could be the incorporation of nanocarriers with an improved drug adsorption uptake and controlled release, provided that the incorporated guest structures does not alter the mechanical properties of the insert, while the porous structure of the nanofiller remains fully accessible in the mixed formulation. [18–20] Among the potential candidates, high-surface area porous materials such as metal-organic frameworks (MOFs) provide an avenue to achieve these requirements.[21] MOFs are crystalline materials formed by the union of metal centres and organic likers. The self-assembly of metal clusters (or nodes) and organic ligands allows the design of a large number of 1D to 3D networks characterized by high surface area, a large pore volume and tuneable host-guest interactions.[22] Over the last few years, these materials have shown promise as a potential platform for drug delivery in powder form.[23,24] Recent studies from Gandara-Loe *et al.* have shown that MOFs can store a large amount of brimonidine tartrate (up to 600 mg of drug per gram of MOF), and with an extended release time of up to 12 days, in the specific case of UiO-67. Furthermore, *in vitro* cytotoxicity assays have demonstrated the low toxicity of UiO-67 for retinal photoreceptor cells.[25] The excellent performance of UiO-67 is motivated by the presence of large tetrahedral and octahedral cages in the micro/mesoporous range.[26] Taking into account these excellent properties, the successful incorporation of these 3D porous networks in continuous polymeric matrices will offer a new perspective in nanomedicine with more suitable nanocomposite materials (instead of working with powders), with novel functionalities (e.g., drug delivery properties), to be used either as micro-inserts (e.g. punctal plug in lacrimal or sub-tenon cavities) or as a component in contact lens. [27,28]

Polymer-MOF nanocomposite materials have already been reported in the literature as potential candidates for gas adsorption/separation processes such as, $CO_2/N_2$ or $CO_2/CH_4$ separation or ethylene adsorption.[29,30] There are recent studies on the use of HKUST-1/polyurethane nanocomposite membranes for drug encapsulation and controlled release.[31] However, the understanding of molecular accessibility in liquid phase adsorption processes is still a challenge due to the different nature of the polymeric network and the MOF nanofiller. To the best of our knowledge, polymer-MOFs nanocomposite films have not yet been tested as a drug delivery carrier for ocular therapeutics.

Based on these premises, the main goal of this work is to report an optimal synthesis of functional MOF-based polyurethane thin films, and to evaluate the performance of these UiO-67@PU nanocomposites for brimonidine adsorption/release in liquid phase. The successful development of these functional materials (MOF@polymer) will open the gate towards the application of these devices in a number of ocular disorders that require a controlled and prolonged release of drugs, from glaucoma treatment to post-surgical treatments by anti-inflammatory drugs.

## 2. Experimental section

### 2.1. UiO-67 synthesis

UiO-67 was synthetized based on the procedure reported in the literature by Katz et al.[32] Briefly, 0.268 g of $ZrCl_4$ were dissolved in a mixture of 20 mL of N,N-dimethylformamide (DMF) and 2 mL of concentrated HCl. In a second vessel 0.360 g of 4,4'-biphenyldicarboxylic acid (BDPC) were dissolved in 40 mL of DMF. The two solutions were mixed and maintained under sonication for 30 min. The final solution was transferred to a 200 mL glass jar, closed tightly and kept at 80ºC overnight. The resulting white solid was filtered and washed first with DMF (2x30 mL) and then with ethanol (2x30 mL). The sample was activated first under low vacuum conditions (13×10$^{-3}$ Pa) up to 90ºC and, afterwards, at 150ºC for 3 h under ultra-high vacuum conditions.

### 2.2. UiO-67@PU synthesis

The UiO-67@PU nanocomposite films were fabricated by following the procedures described below. Polyurethane (PU) solution was prepared by dissolving poly [4,4'-methylenebis (phenyl isocyanate)-alt-1,4-butanediol/di(propylene glycol)/polyurethane] pellets (purchased from Sigma Aldrich and used without further alterations) in tetrahydrofuran (THF) for 24-48 hours until complete dissolution of the polymer pellets. 30 wt.% UiO-67@PU nanocomposites and pristine PU films were produced by the dispersion of a specified amount of previously synthesized MOF particles (of a required wt.%) in a small amount of THF (930 mg of MOF per 1 mL of THF) before their incorporation into the PU-THF solution. The dispersion was performed by a combination of sonication (5 min) and magnetic stirring (20 min, 80 rpm). This strategy, followed by Cohen *et al.* [33], has proven to be a versatile approach for the preparation of homogeneous polymer-MOF nanocomposites. The thin films were subsequently cast onto a glass substrate *via* the doctor blade technique using a casting speed of 10 mm/s to achieve membranes of ~50 μm in thickness.[30,34]

### 2.3. Synchrotron X-ray powder diffraction (SXRPD) analysis

Synchrotron X-ray powder diffraction data (SXRPD) were collected on the powder diffraction end station of the MSPD beamline at synchrotron ALBA in Spain, using a MYTHEN detector and a wavelength of 0.4227 Å. The experiments were performed in an *ad hoc* capillary reaction cell (fused silica capillary, inner diameter 0.7 mm, outer diameter 0.85 mm). SXRPD measurements were performed at 25ºC to the as-synthesized UiO-67, PU and the UiO-67@PU films, and also to the UiO-67@PU films after brimonidine adsorption. The reference spectra for brimonidine tartrate powder was also determined.

### 2.4. Thermogravimetric analysis (TGA)

Thermogravimetric analysis data of UiO-67, PU film and UiO-67@PU film were obtained using TG-DTA METTLER TOLEDO equipment model TG/SDTA851e/SF/1100. The samples were measured using an alumina sample holder and temperature range of 25ºC-600ºC with a heating rate of 5 ºC/min under $N_2$ flow.

## 2.5. Scanning electron microscopy (SEM) evaluation

Cross-section micrographs were recorded using a Hitachi scanning electron microscope model S3000N. This microscope is equipped with Bruker brand X-ray detector (model Xflash 3001) for EDS microanalysis and mapping. Samples were kept under cryogenic conditions (liquid $N_2$) before the analysis in order to obtain a high-quality cross section and avoid surface alterations during the sectioning process.

## 2.6. Nitrogen and ethylene adsorption/desorption isotherms

Textural properties and gas phase accessibility of the different samples were evaluated by gas physisorption, i.e. nitrogen adsorption at -196ºC and ethylene adsorption at 25ºC. Nitrogen gas adsorption measurements were performed in a homemade fully automated manometric equipment designed and constructed by the Advanced Materials Group (LMA), now commercialized as N2GSorb-6 (Gas to Materials Technologies, www.g2mtech.com). Nitrogen adsorption data were used to calculate a) the total pore volume ($V_t$) at relative pressure of 0.95, b) the BET surface area ($S_{BET}$) and c) the micropore volume ($V_{N2}$), after application of the Dubinin-Radushkevich (DR) equation. Ethylene adsorption experiments were performed in a home-built fully automated manometric equipment, now commercialized by Quantachrome Corp. as VSTAR. Before the experiments, samples were degassed at 100ºC for 8 h under high vacuum conditions ($10^{-5}$ torr).

## 2.7. Loading and release experiments

Brimonidine tartrate quantification was done based on the High Performance Liquid Chromatography method developed by Karamanos *et al*.[35] A stock solution of 1500 ppm of brimonidine tartrate was prepared by dissolving 1.5 g in 1000 mL of ultrapure water. Calibration curve was constructed by measuring concentrations from 2 to 15 ppm using chromatographic conditions, analytical column Supelcosil LC-18, 5 µm, 250 x 4.6 mm i.d. stainless steel (Supelco, Bellfonte, PA, USA) equipped with RP-18 precolumn, 20 x 4.6 mm i.d. (Supelco). The mobile phase was a mixture 9:1 (v/v) of 10 mM triethylamine pH 3.2 buffer and acetonitrile. The separation was performed at room temperature, at a flow rate of 1.0 mL/min, injection volume of 20 µL, and the detection of the brimonidine at 248 nm.

*2.7.1. Brimonidine loading experiments*

Brimonidine adsorption isotherms were performed at 25ºC using a group of aqueous solution (pH = 7) prepared from the stock solution with an initial concentration of 250 ppm, 500 ppm, 750 ppm, 1000 ppm and 1500 ppm of brimonidine tartrate. The nanocomposite films were degassed at 100ºC overnight before the experiment. Approximately 100 mg of film were placed in contact with 50 mL of solution at each of the concentrations described above and left under stirring until equilibrium was reached. Aliquots were taken at different time intervals in order to evaluate the adsorption kinetics of the films.

The quantification of brimonidine was determined using High Performance Liquid Chromatography (HPLC) by diluting each aliquot 1:100 and using the method described above.

*2.7.2. Brimonidine release experiments*

100 mg of UiO-67@PU film, previously degassed, was loaded with brimonidine by contacting it with 50 mL of a 1500 ppm brimonidine tartrate aqueous solution. The system was left at 25ºC under stirring for 24 h to ensure full equilibrium. After this time the film was separated from the solution and an aliquot was taken to determine the maximum loading amount. The brimonidine-loaded film was washed several times with ultrapure water and dried under vacuum at 60ºC for 6 h. The dried brimonidine loaded film was immersed in 50 mL of physiological solution (PBS) and aliquots were taken at different times up to 14 days. The aliquots were diluted 1:100 and brimonidine quantification was performed using the HPLC method described above.

## 3. Results and discussion

*3.1. Characterization of the synthesized films and accessibility of the embedded MOFs*

The crystallinity of the synthesized materials has been evaluated through synchrotron X-ray powder diffraction measurements (SXRPD). **Figure 1** shows the comparative SXRPD patterns for the as-synthesized UiO-67 crystals, obtained by solvothermal method, and the UiO-67@PU film. The SXRPD pattern of the UiO-67 sample perfectly fits with the simulated pattern and with those previously described in the literature, thus confirming the quality and reproducibility of the synthetized MOF.[32] Concerning the UiO-67@PU nanocomposite material, the SXRPD pattern confirms the presence of a semi-crystalline system, with the combination of crystallinity due to UiO-67 nanoparticles and the amorphous background from the PU matrix. The PU matrix is characterized by a broad peak between $2\theta$ = 6º – 10º (see **Figure S1**), whereas the main diffraction peaks of the MOF can be clearly appreciated at $2\theta$ = 2.3º-2.6º. These results confirm the preservation of the 3D network in the UiO-67 nanocrystals upon incorporation in the polymeric matrix, and their excellent crystallinity.

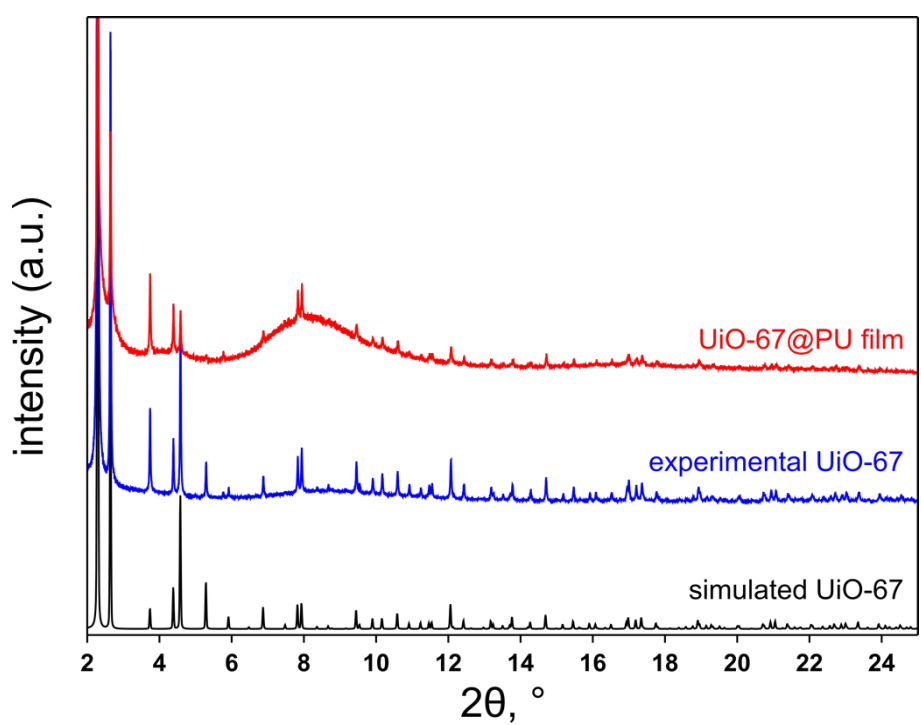

**Figure 1**. Synchrotron XRPD experimental patterns of UiO-67 and UiO-67@PU film accompanied by simulated pattern of UiO-67.

Morphologically, UiO-67@PU film is a semi-transparent and flexible composite material (**Figure S2**) with high versatility for the production of different ocular devices. As it is shown in **Figure 2,** the nanocomposite film is formed by MOF nanocrystals (average crystal size 100-200 nm) embedded into the polyurethane matrix, giving a film of approximately 50 µm thick. **Figure 2c** shows the relatively uniform distribution of the

UiO-67 nanocrystals within the PU matrix, an observation that was further confirmed by specific Zr-mapping experiments (**Figure 2d**). Previous results described in the literature for gas separation using similar composites have anticipated that the accessibility (permeation of gases) decreases with the thickness of the film.[34,36] Based on this assumption and taking into account the objectives of this study (liquid phase adsorption processes usually possess lower kinetics compared to gas adsorption processes), we assume that a film of 50 µm can be considered as a good approach. Furthermore, 30 wt.% MOF loading can be considered as an upper limit to keep a good balance between thermomechanical and toughness properties for a potential future application.[34,37]

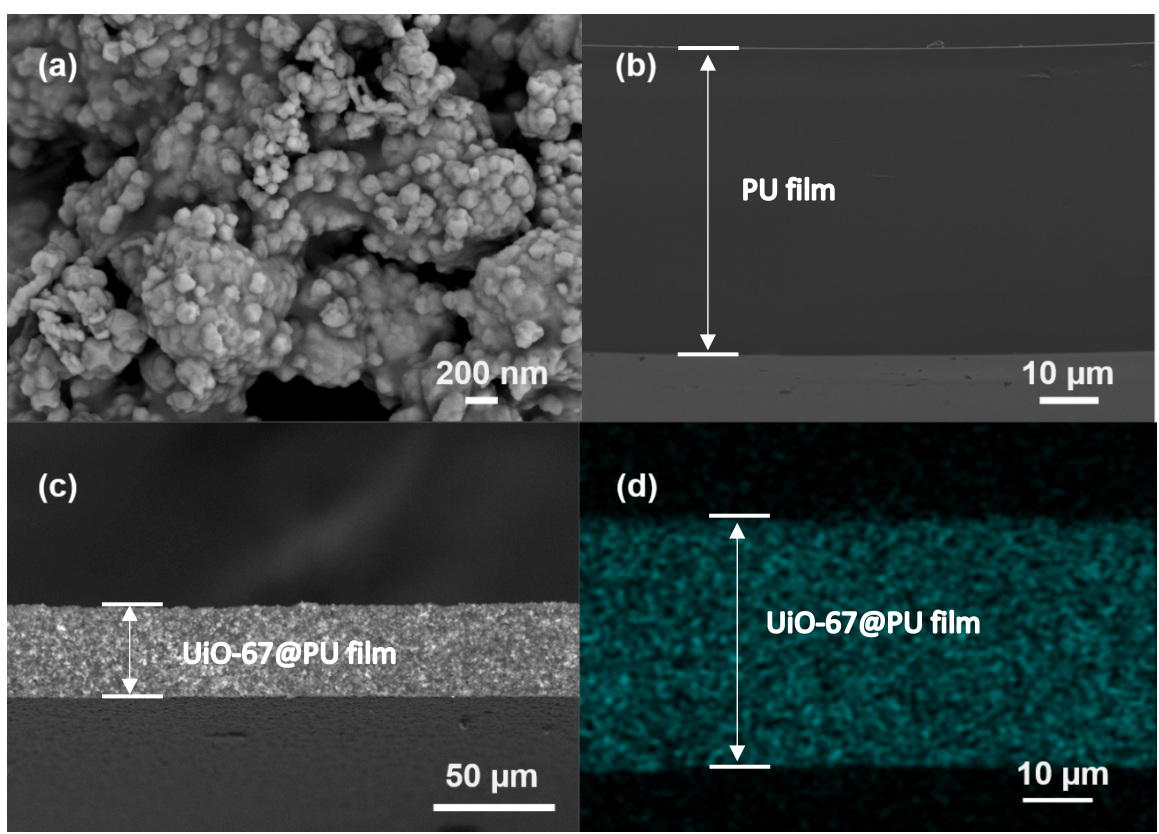

**Figure 2.** SEM micrograph of (a) as-synthesized UiO-67 nanocrystals, (b) cross-section of a 50 µm thick neat PU film, (c) cross-section of UiO-67@PU 50-µm film and (d) Zr EDX mapping (green colour) of a cross-section of UiO-67@PU nanocomposite film.

Thermogravimetric (TGA) analyses were used to evaluate the thermal stability of the nanocomposite film compared to the pure components (PU and UiO-67). Polyurethane and UiO-67 nanoparticles exhibit characteristic decomposition profiles with very sharp and symmetric decompositions peaks, as appreciated in **Figure 3**. For instance, the pure

PU film exhibits a decomposition profile with a well-defined decomposition peak centred at 337ºC and a small shoulder at 430ºC, which is typical of polyurethane materials.[38] In the case of UiO-67, the TGA profile shows the release of the solvent at 135ºC and the main framework decomposition close to 550 ºC.[26] **Figure 3** also shows the TGA profile for the UiO-67@PU nanocomposite film. In this case the scenario is more complex. As it can be appreciated, the nanocomposite material exhibits a broad decomposition profile with a main peak located in between 200 ºC and 300 ºC. Interestingly, this peak is not symmetric and clear shoulders can be appreciated at around 217ºC and 278ºC, in addition to the main contribution at 252ºC. Taking into account that 70 wt.% of the composite corresponds to PU, the main contribution at 252ºC must be attributed to the decomposition of the polymeric matrix. Compared to the pure polymer (*ca.* 337ºC), these results indicate a clear shift to lower temperatures upon incorporation of the MOF nanofillers, in close agreement with previous studies reported in the literature.[34] Apparently, the incorporation of the MOF nanocrystals limits the cross-linking between PU molecular chains, thus reducing their thermal stability. For the sake of clarity, a deconvolution of the DTGA profile for the nanocomposite system can be seen in **Figure S3**. In addition to the decomposition of the polymeric matrix, the aforementioned shoulders must be attributed to solvent removal (*ca*. 217ºC) and to the secondary contribution in the decomposition of the PU matrix (*ca.* 278ºC). Furthermore, the nanocomposite material exhibits an additional decomposition peak at 528ºC, unambiguously attributed to the degradation of the embedded MOF. This finding constitutes another proof about the successful incorporation of the MOF crystals in the polymeric matrix. Table S1 contains a summary of the TGA results for the three samples evaluated.

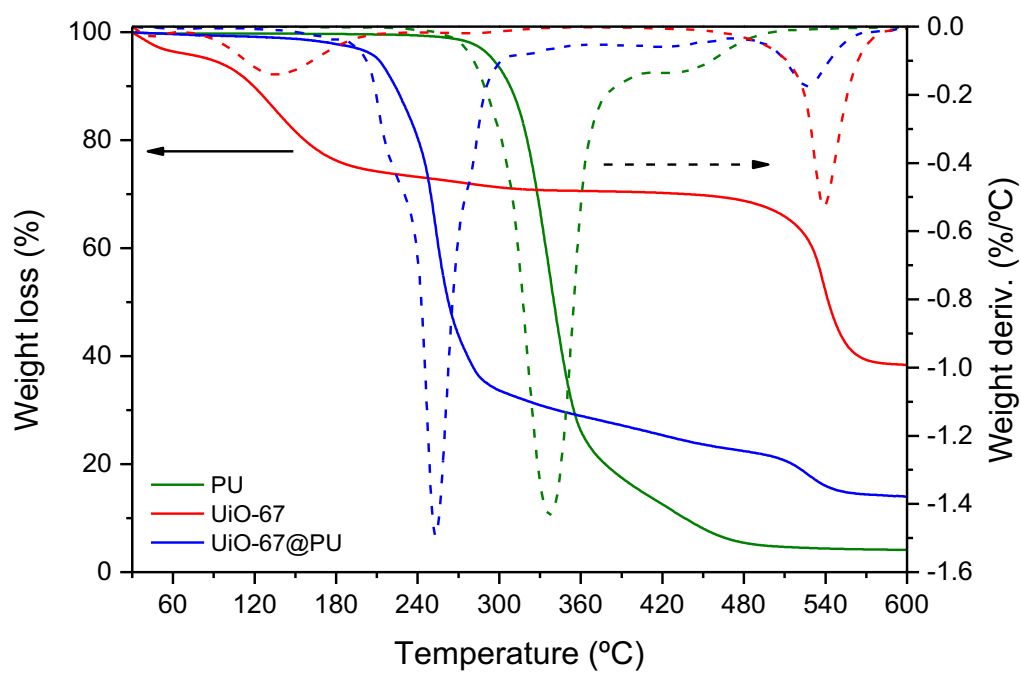

**Figure 3.** Thermogravimetric analysis (TGA and DTGA) of PU, UiO-67 and UiO-67@PU film.

To check the accessibility of the 3D porous network in UiO-67@PU nanocomposite films to gas molecules, the nitrogen adsorption/desorption isotherm was performed at -196ºC and compared to that of the pure MOF. As it can be appreciated in **Figure S4**, UiO-67 presents the typical adsorption-desorption isotherm already described elsewhere,[32] with a large uptake at low relative pressures due to its highly microporous framework, and the associated step at $p/p_0$ ~ 0.15 attributed to the presence of wider pores (small mesopores). This observation is in close agreement with the presence of two kind of cavities in UiO-67, tetrahedral and octahedral cages with a diameter of 1.1 and 2.3 nm, respectively.[32] Interestingly, in the specific case of the UiO-67@PU film the accessibility for nitrogen at cryogenic temperatures is completely suppressed over the whole relative pressure range evaluated. This observation is in close agreement with previous studies described in the literature for ZIF-8 and ZIF-7 loaded polymeric matrices.[37] Apparently, nitrogen with a quadrupolar moment is not able to diffuse through the rubbery polymeric network at cryogenic temperatures. Despite the inaccessibility of nitrogen to the embedded MOF crystals, this observation does not necessarily reflect the real scenario in the composite material. Based on our previous experience, adsorption of non-polar molecules (for instance, hydrocarbons) constitutes a complementary tool to evaluate the porous structure in these MOF@polymer nanocomposites. **Figure 4** shows the ethylene adsorption/desorption isotherms at 25ºC for the pure PU, UiO-67 and the nanocomposite. These results show that, contrary to $N_2$, ethylene is indeed able to access the inner porous structure in this kind of materials. Whereas the pure PU film exhibits an adsorption capacity close to 0 mmol/g, UiO-67 nanoparticles are able to adsorb up to 1.31 mmol/g at a pressure of 1 bar. For the UiO-67@PU nanocomposite sample, the total adsorption capacity for ethylene at 1 bar is *ca.* 0.18 mmol/g. After a normalization to the total amount of MOF (considering that the composite contains *ca.* 30 wt.%), this value scales up to a total uptake of 0.59 mmol/$g_{MOF}$. Compared to the pure UiO-67, this result constitutes a reduction of 55% in the adsorption capacity of the embedded crystals, i.e. embedded nanocrystals are indeed accessible to gas molecules, although only partially.

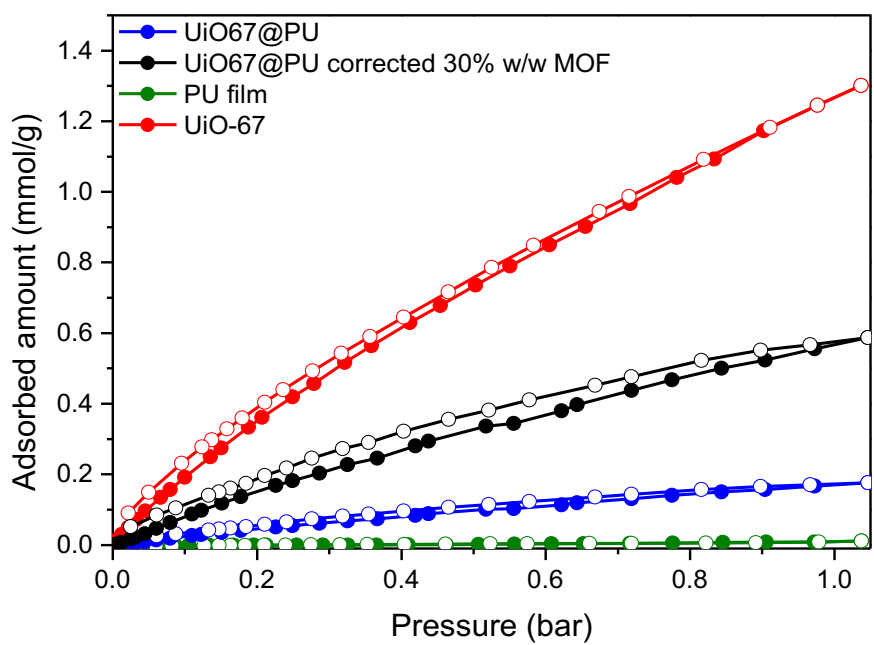

**Figure 4.** Ethylene adsorption (filled symbol)-desorption (open symbol) isotherms at 25ºC in as-synthesized UiO-67, PU and UiO-67@PU films.

### 3.2. Brimonidine adsorption and release

Brimonidine adsorption isotherms were performed in aqueous media (ultrapure water) and at room temperature in order to quantify the maximum amount of drug adsorbed in the porous structure of the synthesized films. As is shown in **Figure 5,** while the adsorption in the pure PU film is close to 0 mg/g, the maximum brimonidine adsorption capacity in the UiO-67@PU film (at an equilibrium time of 4h; see **Figure S5**) obtained from the Langmuir model achieves a value of 58.4 mg of brimonidine per gram of film, i.e. 194.7 mg of brimonidine per gram of UiO-67 (considering the nominal value of 30 wt.% of UiO-67 in the film). This value differs from that reported in the literature for pure UiO-67 nanoparticles (*ca.* 600 mg$_{brimonidine}$/g$_{MOF}$).[25] The reduction in the adsorption capacity for the nanocomposite (around 67% reduction) is in close agreement with the gas-phase ethylene adsorption measurements described above (ethylene was able to access 45% of the porosity whereas brimonidine only 32.5% of the MOF porous network). Although these numbers must be optimized, this finding constitutes an important development elucidating the potential application of these MOF-doped polymeric matrices for liquid-phase adsorption/desorption processes. Even though these processes are performed in the presence of a solvent (for instance, an aqueous solution), the embedded MOFs are able to preserve a similar accessibility to the target molecule (e.g. ocular drug), compared to similar measurements in gas phase, i.e. in the absence of solvent. These results suggest that UiO-67 cavities are able to host both ethylene (molecular size of 4.7 x 9.8 Å) and brimonidine (3.28 x 4.18 x 4.84 Å) in a similar extend.[39,40] Compared to the neat PU polymer, the incorporation of UiO-67 nanofillers gives rise to a 60-fold increase in the adsorption capacity for brimonidine tartrate.

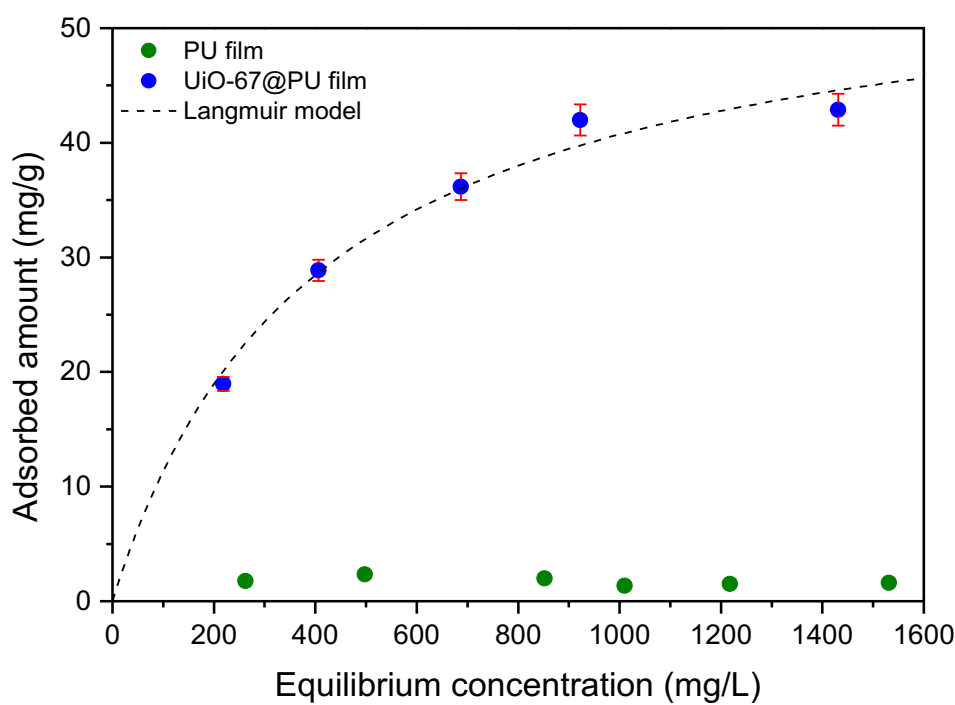

**Figure 5.** Brimonidine liquid-phase adsorption isotherms in PU and UiO-67@PU films at 25ºC ($C_0$ = 1500 ppm).

In order to mimic a potential application in human body, brimonidine release isotherms were performed in PBS solution at room temperature and neutral pH, **Figure 6**. As it can be observed, the UiO-67@PU nanocomposite exhibits a fast release (up to 7 % of the total uptake) in the first minute of the experiment. Afterwards, there is a continuous release with time up to a maximum of 10% of the total brimonidine retained after 14 days exposure. The large release in the first few hours must be attributed to brimonidine weakly interacting with the nanocomposite and/or adsorbed in the external layers/pores of the film. Considering the traditional topical administration of brimonidine, i.e. a patient must take one droplet of brimonidine solution of 2 mg/mL (Alphagan P®, Allergan) every 8h, this means 0.3 mg of brimonidine per day or 4.2 mg in 14 days.[7] Taking into account the total uptake of 58.4 mg/g for our composite, a release of 10 % (5.8 mg/g) after 14 days is within the needs of a normal patient with glaucoma, thus validating our approach. At this point it is important to highlight that we cannot exclude the possibility that some brimonidine is already removed/released from the loaded film during the washing step performed after the loading and before the release experiment (this washing step was designed to remove exclusively the brimonidine retained in the external surface of the film).

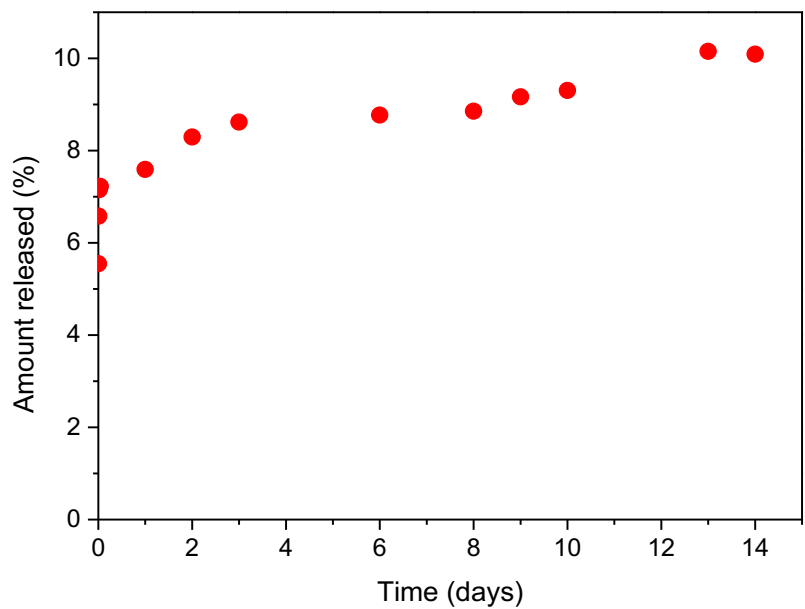

**Figure 6.** Brimonidine tartrate release kinetics at 25ºC in physiological media PBS (loading concentration 1500 ppm)

At this point, the open questions remain the compatibility of the drug with the composite, the stability of the MOF structure after the loading process and finally, the potential location of the drug molecule in the composite system. Next sections are devoted to answer all these questions.

*3.3. Brimonidine-composite compatibility and stability studies*

Structural stability of the MOF framework is an important parameter to be considered in liquid-phase adsorption processes. It is widely accepted in the literature that MOF materials can exhibit a limited stability in the presence of aqueous environments or after the incorporation of the drug.[41] In the specific case of UiO-67, it is well-known that upon exposure to water or moisture, this system exhibits a large instability due to the hydrolysis of the linker-metal bonds, and the associated pore collapse.[42–44] However, the partial amorphization of the UiO-67 nanoparticles during the adsorption/release of brimonidine has been very useful to extend the released kinetics beyond 12 days, as described before by some of us.[25]

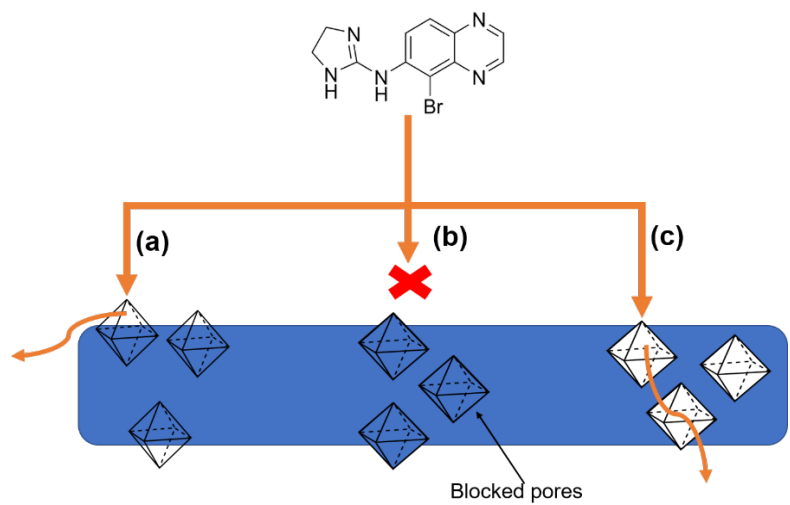

**Figure 7.** Scheme of possible scenarios for Brimonidine adsorption in MOF@polymer composites: (a) adsorption in the peripheric MOF crystals, (b) fully inaccessible and (c) fully accessible embedded MOF nanocrystals.

In addition to the structural stability, another concern is the adsorption mechanism. Adsorption of brimonidine into MOF-based polymeric films can be explained via three potential scenarios. As summarized in **Figure 7**, brimonidine can be adsorbed only in those MOF crystals located in the periphery of the PU film (option A), brimonidine can be adsorbed only in the polymeric matrix, i.e. MOF nanocrystals are completely blocked (option B) or it can be adsorbed equally in the different crystals homogenously distributed within the PU film (option C). To identify which of these options is the most plausible to explain the adsorption mechanism, the UiO-67@PU nanocomposite has been thoroughly evaluated before and after adsorption of brimonidine using synchrotron X-ray diffraction, thermogravimetry (TGA) and FTIR.

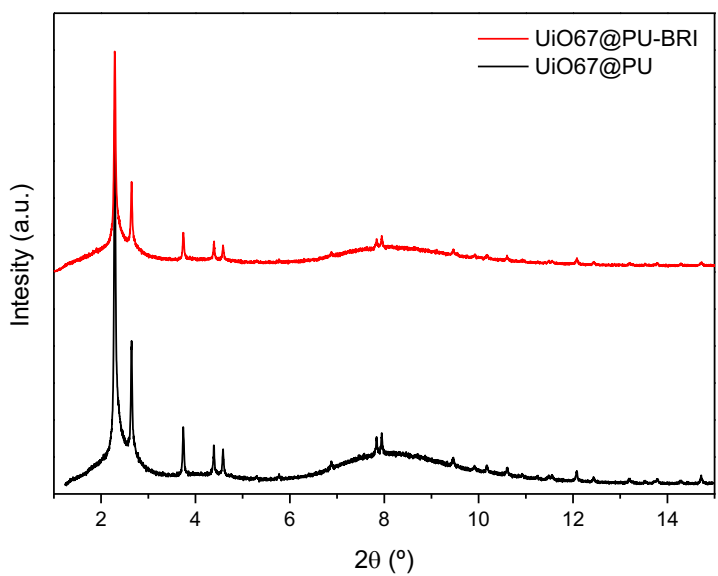

**Figure 8.** Synchrotron X-ray powder diffraction patterns of UiO-67@PU film before and after being exposed to the brimonidine solution.

Synchrotron X-ray diffraction measurements were performed in order to elucidate the structural parameters of the UiO-67 embedded crystals before and after the loading of brimonidine. As it can be observed in **Figure 8** both patterns are rather similar even after exposure to the brimonidine aqueous solution for several days. These results are contrary to the performance of the pure MOF (**Figure S6**), where a significant structural deterioration was identified after 1 day in contact with water and confirms the improved structural stability of UiO-67 upon encapsulation in the PU matrix. [25]

Although the cavities in UiO-67 (octahedral of 2.3 nm and tetrahedral of 1.15 nm) are large enough to accommodate the brimonidine molecule, the open question at this point is how to ascertain if brimonidine is able to take advantage of these cavities.[32] Synchrotron X-ray diffraction measurements of the pure brimonidine tartrate (**Figure S7**) show a rich XRD pattern with a large number of peaks in the range 2 and 18º, confirming the high crystallinity of this molecule. The absence of these peaks in the SXRPD pattern of the brimonidine-loaded UiO-67@PU nanocomposite (**Figure 8**) could be *a priori* an evidence of the absence of brimonidine both in the polymeric network and in the embedded MOF nanocrystals. However, this observation would be in contradiction with brimonidine adsorption measurements reported in **Figure 5**. This inconsistency must be explained due to the amorphization of the drug upon adsorption, thus explaining the absence of peaks in the SXRPD pattern. This hypothesis would be in agreement with the encapsulation of the drug in the MOF cavities, with the associated limitation for these molecules to arrange in a periodic fashion. These conclusions are also supported by previous studies dealing with the adsorption/release of brimonidine through ocular devices suggesting the transformation of crystalline brimonidine into an amorphous phase once it is adsorbed into the material.[45–47]

The unit cell parameters deduced for the embedded UiO-67 crystals after Rietveld refinement are summarized in **Table 1.** Pure UiO-67 crystals have a cubic unit cell with lattice parameters $a = b = c = 26.8447(9)$ Å. As it can be observed, the lattice parameters remain rather similar after incorporation of the UiO-67 crystals in the polymeric matrix, in close agreement with the high quality of the crystals described in **Figure 1**. Interestingly, lattice parameters do not change after exposure of the UiO-67@PU nanocomposite to an aqueous solution of brimonidine. Although these results confirm the large stability of UiO-67 nanocrystals in an aqueous environment upon incorporation in the PU matrix, these are not conclusive about the location of brimonidine upon adsorption. Unfortunately, Rietveld refinement analysis of the embedded crystals does not allow to answer this question due to the limited quality of the SXRPD pattern.

**Table 1.** Summary of structural parameters and adsorption performance of UiO-67, and UiO-67@PU film before and after loading with brimonidine.

|  | UiO-67 | UiO-67@PU | UiO-67@PU-BRI |
| --- | --- | --- | --- |
| Cell parameter $a$ (Å) | 26.8447(9) | 26.8306(6) | 26.8252(6) |
| $S_{BET}$ (cc STP/g) | 3352 | 0 | ---- |
| Ethylene adsorption (mmol/g) | 1.31 | 0.18 | ---- |
| Brimonidine adsorption (mg/g) | 600[25] | 58.4 | ---- |
| Brimonidine release (%) | 50% (12 days) | 10% (14 days) | ---- |

To further ascertain the adsorption mechanism, TGA analysis was performed in the UiO-67@PU film after the loading of brimonidine. For clarity, the TGA of pure brimonidine tartrate has been included in **Figure S8**. Brimonidine tartrate exhibits a single decomposition peak at around 210 ºC. A closer look to the TGA profile for the brimonidine-loaded UiO-67@PU nanocomposite (**Figure S9**) shows that the TGA peaks corresponding to the decomposition of the PU matrix and the UiO-67 crystals are shifted to higher temperatures upon adsorption. In addition, the thermogram shows an additional tiny peak at 210ºC, not present in the un-loaded UiO-67@PU material, that can be attributed to brimonidine within the composite film (blue peak deconvoluted in **Figure S10**). Although the shifts observed in **Figure S9** for the decomposition of PU and UiO-67 upon brimonidine adsorption could be an indication of the presence of brimonidine in both domains, the real location of the drug remains still an open question. Last but not least, it is important to highlight that the quantification of the tinny peak at 210ºC corresponds to ~ 23 mg $_{Brimonidine}$/g$_{composite\ film}$. Although this is a rough estimation, we cannot exclude that around 40% of the brimonidine loaded at 1500 ppm (**Figure 5**) could be lost during the washing step applied before the TGA analysis. A similar hypothesis could be used to explain the low release achieved in **Figure 6**.

Finally, the presence of the drug has been evaluated using FTIR of the UiO-67@PU film before and after loading with brimonidine, **Figure 9**. The FTIR spectra for the individual components have also been included for clarity. As it can be observed, before loading, the FTIR spectra of the UiO-67@PU film shows the characteristic peaks of PU and UiO-67. PU has a characteristic peak at 3329 cm$^{-1}$ attributed to the stretching of the NH bond (**Figure 9a**). In addition there are two contributions at 1724 cm$^{-1}$ and 1696 cm$^{-1}$ due to the poly(caprolactone) ester bond, and the -CH stretching vibration at 2944 cm$^{-1}$, among others.[48,49] The characteristic peaks of UiO-67 can be observed at 1594 cm$^{-1}$, 1528 cm$^{-1}$ and 1411 cm$^{-1}$ due to the stretching vibrations of the carboxylate group of the ligands and, the peaks at 815 cm$^{-1}$, 766 cm$^{-1}$ and 652 cm$^{-1}$ due to the Zr-O stretching vibrations.[50,51]

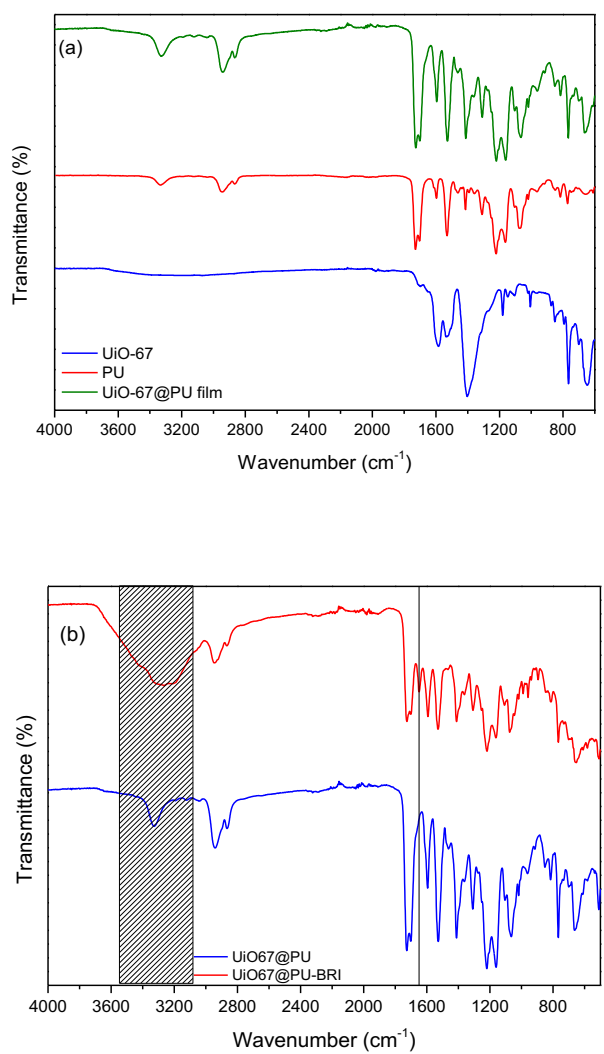

**Figure 9.** FTIR spectra of (a) UiO-67 (bottom), PU (middle) and UiO67@PU film (upper), and (b) UiO67@PU film before (bottom) and after (upper) loading with brimonidine.

As already reported in the literature, brimonidine tartrate also presents characteristic vibrations in the IR range. These characteristic vibrations include peaks at 3212 and 3268 cm$^{-1}$ owing to N-H stretching vibration from the secondary amine groups (RR'-NH). Peaks around 1650 cm$^{-1}$ are attributed to C=O stretching and -CN stretching appears at 1284 cm$^{-1}$.[52–54]

The most remarkable feature of the FTIR spectra of UiO-67@PU after loading brimonidine is, in addition to the bands described above due to PU and UiO-67, the presence of a wide contribution around 3575-3074 cm$^{-1}$. This broad contribution could be associated to the overlapping of signals from adsorbed H$_2$O (O-H stretching at 3404

cm$^{-1}$), and to the stretching -NH vibrations characteristics of urea and urethane bonds (3333 cm$^{-1}$) in PU.[48,49] However, taking into account that the brimonidine-loaded sample has been vacuum dried at 60ºC before the FTIR spectra, and the absence of this wide contribution in the drug-free nanocomposite film, the presence of this broad contribution must be unambiguously attributed to the presence of brimonidine chemically interacting with the composite via hydrogen bonding with surface oxygen and nitrogen groups. This finding is supported by the presence of a new peak at 1650 cm$^{-1}$ (solid line in **Figure 9b**) in the loaded film due to the C=O groups of the brimonidine tartrate. These assignments are in perfect agreement with previous studies in NH$_2$-MIL-88(Fe) loaded with brimonidine.[53] In summary, FTIR spectra clearly confirm the presence of the drug in the UiO-67@PU film, although the real location, either in the polymeric matrix or in the UiO-67 network cannot be easily identified.

*Conclusions*

We have successfully developed a novel UiO-67-based polyurethane film with an excellent adsorption/release performance for an ocular drug such as brimonidine tartrate. Synchrotron X-ray powder diffraction measurements confirms the high quality of the MOF nanocrystals when embedded in a hydrophobic polymer such as PU and their improved stability in an aqueous environment, compared to the pure MOF. Although the inner porous structure is not accessible to nitrogen with a quadrupole moment, this is not the case for the adsorption of non-polar molecules (e.g., hydrocarbons) at room temperature. Although the partial accessibility of the embedded MOFs limits the brimonidine adsorption performance, the UiO-67@PU composite gives rise to a 60-fold improvement compared to the neat PU film. Synchrotron XRPD, TGA and FTIR measurements of the composite before and after loading brimonidine confirm the presence of the drug within the UiO-67@PU film, although the real role of the polymer matrix and the UiO-67 nanocrystals cannot be conclusively confirmed. The total brimonidine uptake of the composite is as high as 58.4 mg$_{BRI}$ per gram of composite or 194.7 mg$_{BRI}$ per gram of MOF. These results in liquid-phase are highly promising and open the door to the design of novel polymeric inserts with functional properties and improved performance (for instance with drug delivery properties), to be applied in a number of ophthalmological disorders, either as a component of contact lens, in the composition of lacrimal stoppers (e.g., punctal plugs) or in sub-tenon inserts.

**Acknowledgments**

The authors would like to acknowledge financial support from the MINECO (MAT2016-80285-p), Spanish ALBA synchrotron (Project 2020014008) and H2020 (MSCA-RISE-2016/NanoMed Project). B. E. S. thanks the Minas Gerais Research Foundation (FAPEMIG CNPJ n21.949.888/0001-83) for a DPhil scholarship award. J. C. T. thanks the EPSRC Grant No. EP/N014960/1 and ERC Consolidator Grant (PROMOFS under the grant agreement 771575) for funding.

**ORCID**

J. Gandara-Loe: 0000-0003-1334-4788

B.E. Zouza: 0000-0001-6315-2149

A. Missyul: 0000-0002-0577-4481

J.-C. Tan: 0000-0002-5770-408X

J. Silvestre-Albero: 0000-0002-0303-0817

SUPPORTING INFORMATION

# MOF-Based Polymeric Nanocomposite Films as Potential Materials for Drug Delivery Devices in Ocular Therapeutics

J. Gandara-Loe,[1] B.E. Souza,[2] A. Missyul,[3] G. Giraldo,[4] J.-C. Tan,[2] J. Silvestre Albero[1,*]

[1]Laboratorio de Materiales Avanzados, Departamento de Química Inorgánica-IUMA, Universidad de Alicante, E-03690 San Vicente del Raspeig, Spain

[2]Multifunctional Materials & Composites (MMC) Laboratory, Department of Engineering Science, University of Oxford, Parks Road, Oxford OX1 3PJ, UK

[3] CELLS-ALBA Synchrotron, E-08290, Cerdanyola del Vallés, Spain

[4] Clínica Clofan, Carrera 48 # 19 A 40, Medellín, Colombia.

* Email: joaquin.silvestre@ua.es

*Table of Contents* *Page*

- *Equations applied in this study.*

- 30 wt. % UiO-67 encapsulated in a polymeric (PU) 50 μm film was prepared using the following equation (1):

$$UiO\text{-}67\ wt.\% = \left(\frac{m_{UiO-67}}{m_{UiO-67}+m_{PU}}\right) x\ 100\% \quad (1)$$

where $m_{UiO-67}$ is the weight of UiO-67 nanoparticles dispersed in THF and $m_{PU}$ is the weight of PU pellets dissolved in THF.

- Maximum amount of brimonidine that can be released (2) and real percentage released (3) was calculated using the following equation:

$$m_{bri-max} = \frac{C_0 - C_{eq}}{m_{film}} \quad (2)$$

$$\%\ released = \frac{C_{eq-released} \times v}{m_{bri-max}} \quad (3)$$

where $m_{bri\text{-}max}$ is the maximum amount of brimonidine adsorbed in a given mass of film ($m_{film}$), $C_0$ is the initial concentration of brimonidine, $C_{eq}$ is the concentration after the adsorption reached the equilibrium, $C_{eq\text{-}released}$ is the concentration measured after a given time during the releasing process in PBS solution and $v$ the volume of PBS.

- *Tables*

**Table S1.** Thermogravimetric results of the different samples evaluated.

| Sample | 1st stage | | 2nd stage | | 3rd stage | | |
|---|---|---|---|---|---|---|---|
| | ΔT (°C) | ΔW (wt.%) | ΔT (°C) | ΔW (wt.%) | ΔT (°C) | ΔW (wt.%) | $T_m$ (°C) |
| UiO-67 | 30-200 | 25.3 | 200-440 | 4.7 | 440-600 | 31.5 | 540 |
| PU | 30-190 | 0.4 | 235-500 | 94.7 | 500-600 | 0.7 | 337 |
| UiO-67@PU | 30-190 | 2.7 | 190-370 | 68.9 | 455-600 | 8.4 | 252 |

ΔT: temperature range of the thermal decomposition.
ΔW: Total weight loss at the main decomposition process
$T_m$: The degradation temperature corresponding to the maximum weight loss rate of DTG curve.

- *Figures*

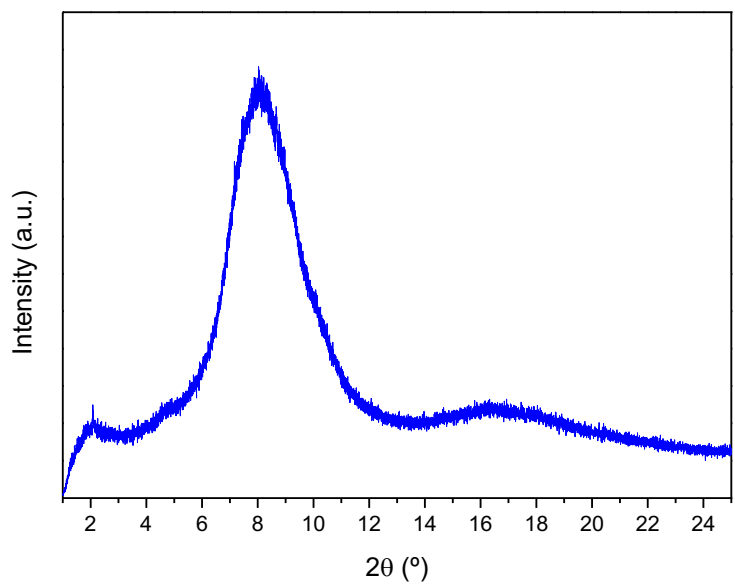

**Figure S1.** Synchrotron X-ray powder diffraction pattern of pure polyurethane film.

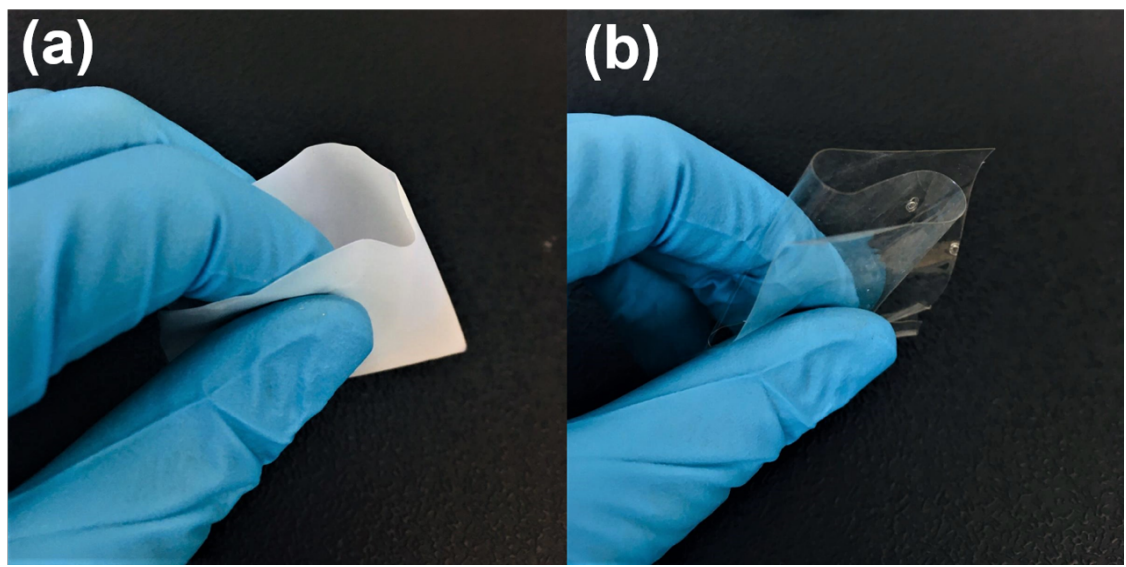

**Figure S2.** Photographs of the different samples prepared by doctor blading, (a) UiO67@PU and (b) PU films.

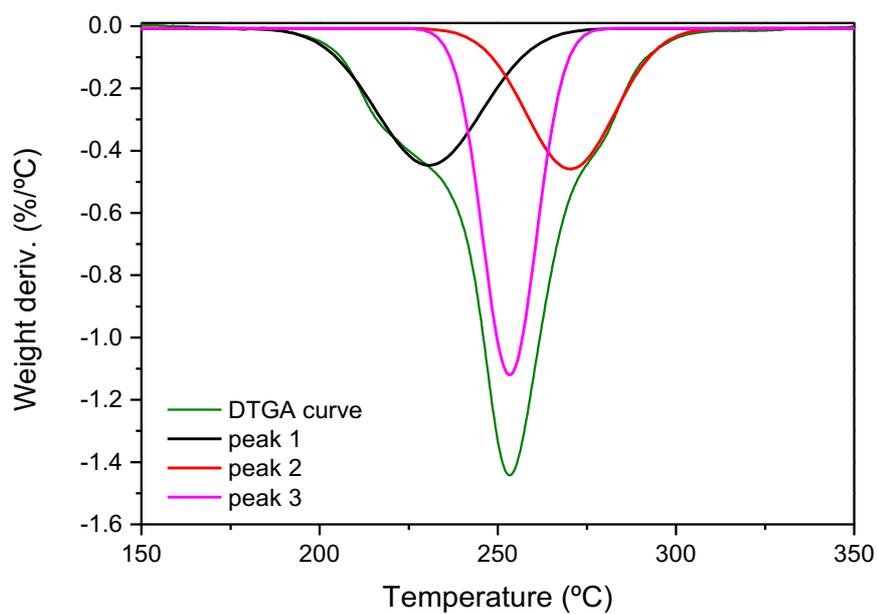

**Figure S3.** Deconvolution of the DTGA profile for UiO-67@PU film.

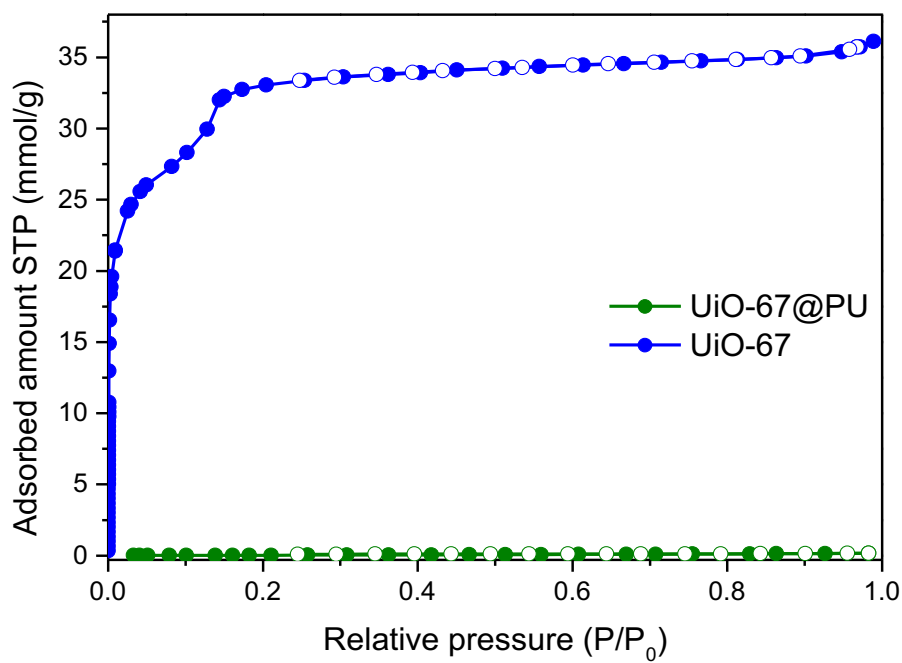

**Figure S4.** Nitrogen adsorption (filled symbols)-desorption (open symbols) isotherms at -196ºC for UiO-67 and UiO-67@PU film.

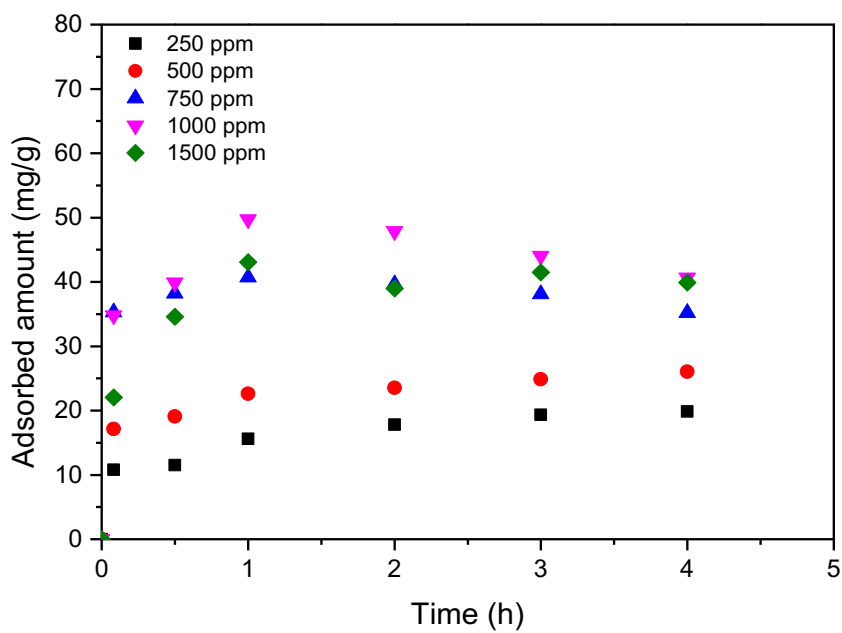

**Figure S5**. Brimonidine adsorption kinetics in the UiO67@PU film at different initial concentrations.

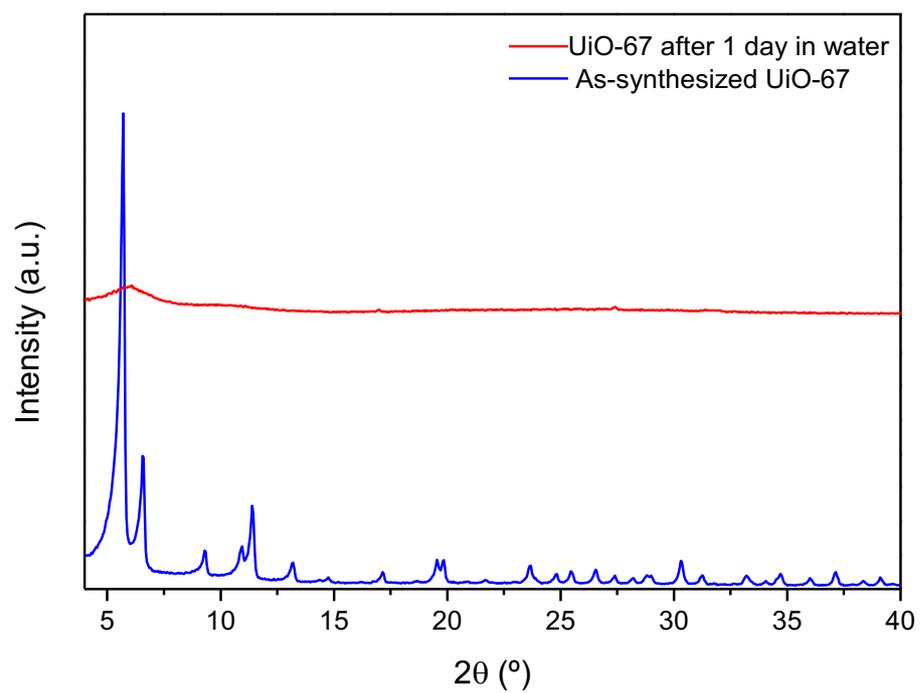

**Figure S6**. X-ray powder diffraction pattern of as-synthesized UiO-67 and after soaking in water for 1 day.[1]

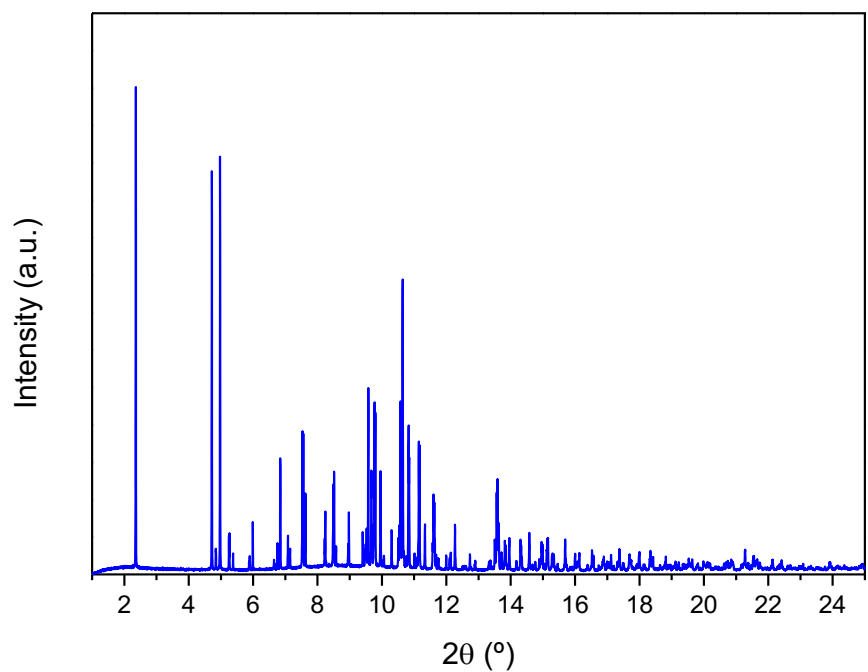

**Figure S7**. Synchrotron X-ray powder diffraction pattern of brimonidine tartrate.

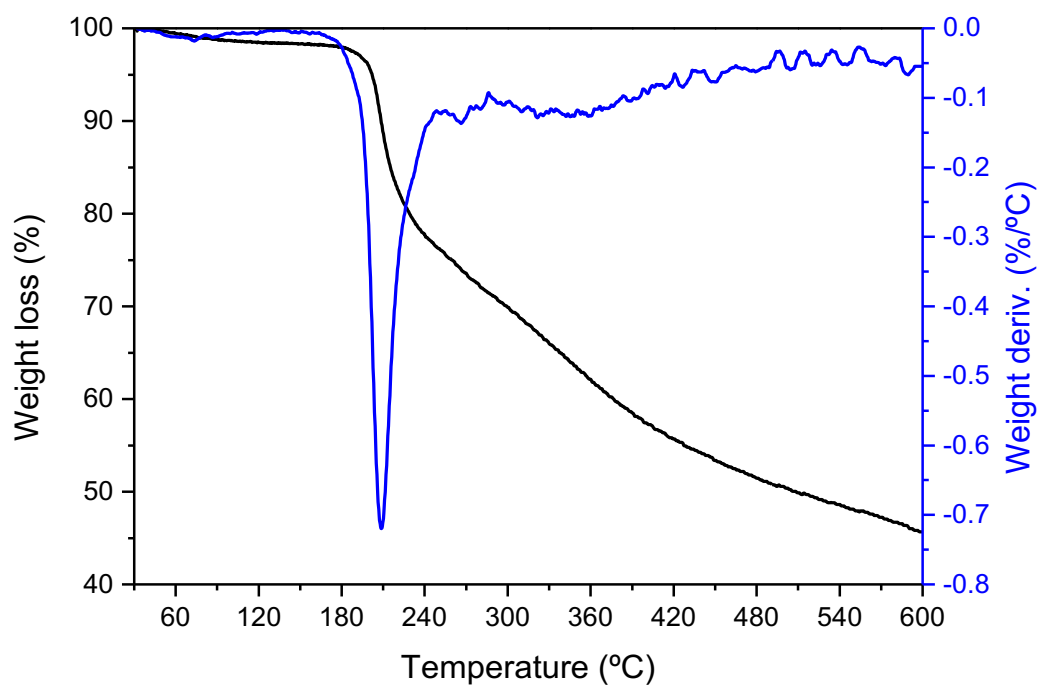

**Figure S8.** TGA-DTGA profiles for brimonidine tartrate.

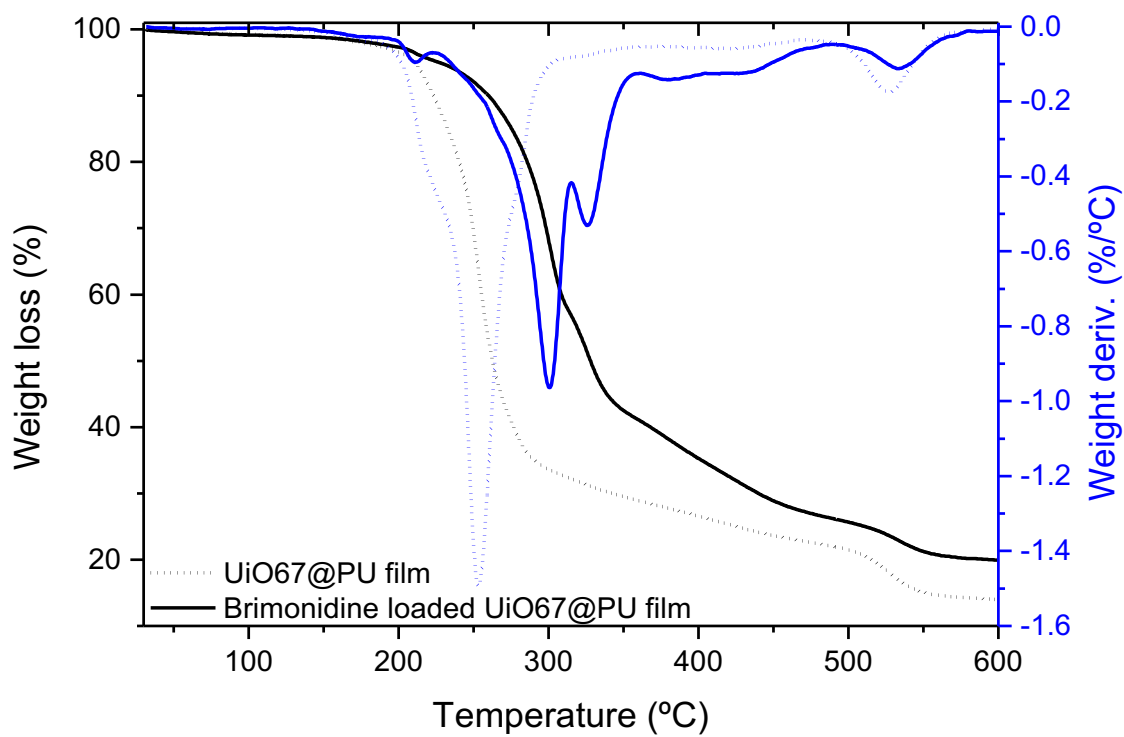

**Figure S9**. TGA-DTGA profiles for UiO-67@PU film before and after loading with brimonidine.

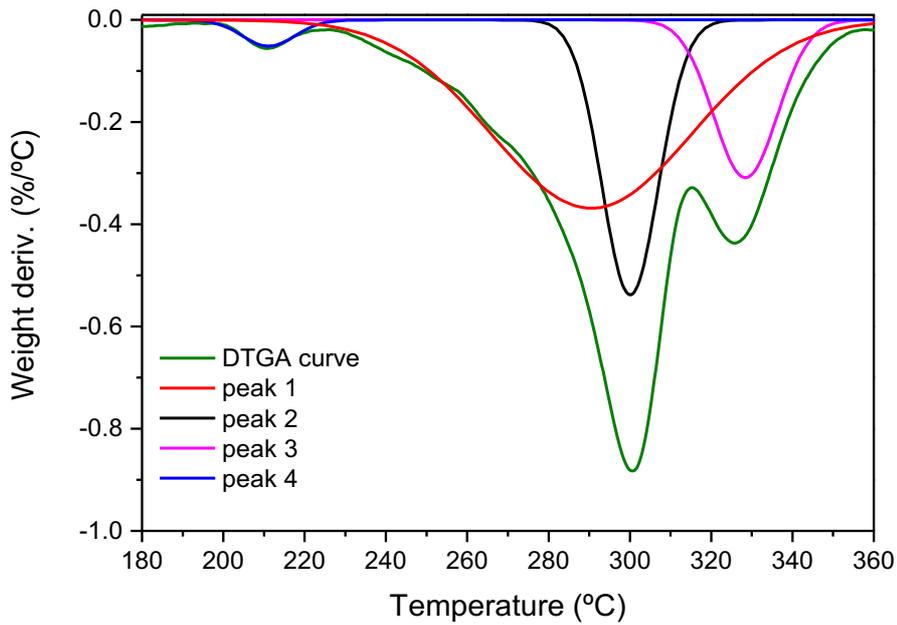

**Figure S10**. Deconvolution of the DTGA profile in brimonidine loaded UiO-67@PU films.

- *Langmuir model for brimonidine adsorption isotherm.*

Adsorption isotherms are defined as the mathematical relationship between the mass of the adsorbed solute per adsorbent mass unit and the solute concentration remained in the solution when the equilibrium has been reached at constant temperature [2]. The most widely used isotherm models for liquid-solid systems are Langmuir, Freundlich and Prausnitz-Radke [3].

Langmuir model (4) was theoretically developed based on the following assumptions: i) the adsorption occurs in specific sites on the surface of the adsorbent, ii) only one molecule is adsorbed in each active site, iii) there are not interactions between adjacent adsorbed molecules and iv) the adsorption heat is the same for all the active sites. This model is mathematically represented as:

$$q_{eq} = \frac{(C_o - C_{eq}) \cdot v}{m} \quad (3)$$

$$q_{eq} = \frac{q_{max} \cdot K \cdot C_{eq}}{1 + K \cdot C_{eq}} \quad (4)$$

Where $C_{eq}$ (mg/L) is the concentration of the solute after the equilibrium has been reached, $q_{eq}$ (mg/g) is the mas of solute adsorbed per unit mass of adsorbent, $q_{max}$ (mg/g) is the maximum amount of solute that can be adsorbed by the adsorbent, $C_o$ is the initial concentration and $K$ (L/mg) is the Langmuir constant related with the heat of adsorption.

The equation was solved using Statistica 10 software of StatSoft Inc by nonlinear estimation with estimation method of Rosenbrook and Quasi-Newton. The values obtained for brimonidine adsorption in UiO-67@PU nanocomposite films are:

$$q_{max} = 58.44 \pm 1.89 \frac{mg}{g}$$

$$K_{Langmuir} = 2.48 \times 10^{-3}$$

- *Brimonidine chromatograph*

Chromatographic conditions used for the quantification of brimonidine were based in the method developed by Karamanos et al [4]. 10 mM triethylamine 3.2 buffer and acetonitrile were used as mobile phase for the column. Figure S10 shows a typical chromatogram for brimonidine quantified by HPLC.

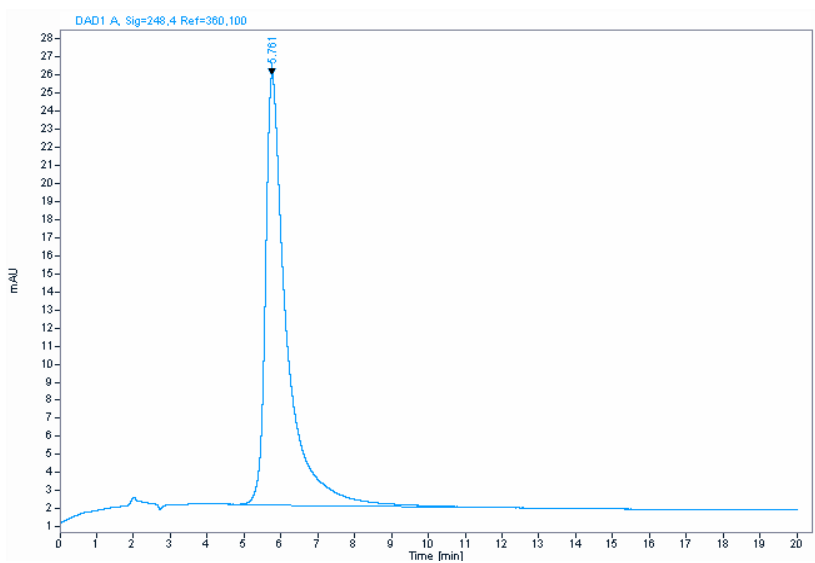

**Figure S11**. Typical chromatogram for brimonidine using HPLC and detected by UV-Vis.

### References

1. Gandara-Loe, J. *et al.* Metal–Organic Frameworks as Drug Delivery Platforms for Ocular Therapeutics. *ACS Appl. Mater. Interfaces* **11**, 1924–1931 (2019).